\documentclass[mathleft,fleqn,%
]{an}
\usepackage{graphicx}
\usepackage[varg]{txfonts}
\usepackage{natbib}
\bibpunct{(}{)}{;}{a}{}{,}
\begin{document}

\Pagespan{1}{}
\Yearpublication{2016}%
\Yearsubmission{2016}%
\Month{0}%
\Volume{999}%
\Issue{0}%
\DOI{asna.201400000}%

\title{Milky Way's Thick and Thin disk: Is there distinct thick disk?}

\author{D. Kawata\inst{1}\fnmsep\thanks{Corresponding author:
        {d.kawata@ucl.ac.uk}}
\and  C. Chiappini\inst{2}
}
\titlerunning{Galactic Disk}
\authorrunning{D. Kawata \& C. Chiappini}
\institute{
 Mullard Space Science Laboratory, University College London, Holmbury St.\ Mary, Dorking, Surrey, RH5 6NT, UK
\and 
 Leibniz-Institut f\"ur Astrophysik Potsdam (AIP), An der Sternwarte 16, 4482 Potsdam, Germany
 }
\received{XXXX}
\accepted{XXXX}
\publonline{XXXX}


\abstract{%
This article is based on our discussion session on Milky Way models at the 592 WE-Heraeus Seminar, Reconstructing the Milky Way's History: Spectroscopic Surveys, Asteroseismology and Chemodynamical models. The discussion focused on the following question:  "Are there distinct thick and thin disks?". The answer to this question depends on the definition one adopts for thin and thick disks. The participants of this discussion converged to the idea that there are at least two different types of disks in the Milky Way. However, there are still important open questions on how to best define these two types of disks (chemically, kinematically, geometrically or by age?). The question of what is the origin of the distinct disks remains open. The future Galactic surveys which are highlighted in this conference should help us answering these questions. The almost one-hour debate involving researchers in the field representing  different modelling approaches (Galactic models such as TRILEGAL, Besan\c con and Galaxia, chemical evolution models, extended distribution functions method, chemodynamics in the cosmological context, and self-consistent cosmological simulations) illustrated how important is to have all these parallel approaches. All approaches have their advantages and shortcomings (also discussed), and different approaches are useful to address specific points that might help us answering the more general question above.
}
\maketitle

\section{Introduction}

Geometrically defined thick disks in addition to a thin disk is ubiquitously found in external galaxies
\citep[e.g.][]{db02}.  
Traditionally, galactic disks are divided into two components by fitting stellar density distributions with thick and thin disk components. The thick disk of the Milky Way was also discovered by the vertical number density profile of stars \citep{yy82,gilr83}. 
Using the photometric data of the Sloan Digital Sky Survey (SDSS) \citealt{jibls08}
analysed the stellar number density map for the Milky Way disk and found that the observed stellar number density distribution in the Milky Way were well fitted by the two (thin and thick) disk components with scale-heights of 300 pc and 900 pc. These are called here the \emph{geometrically} defined thick and thin disks. 

It is a natural assumption that the geometrically thicker disk component is composed by stars possessing higher velocity dispersion and therefore higher peculiar velocities in the solar neighbourhood. High-resolution spectra of \emph{kinematically selected} nearby thick disk stars in the solar neighbourhood showed systematically higher $\alpha$-elements abundance with respect to the iron abundance, [$\alpha$/Fe], compared to kinematically colder thin disk stars with similar [Fe/H] 
\citep[e.g.][]{pncmw00,fbl03}. 
However, the recent studies of a larger number ($\sim$1000) of nearby dwarf stars showed that kinematically selected geometrically thick (thin) disk stars do not always follow the [$\alpha$/Fe]-high ([$\alpha$/Fe]-low) sequence \citep[e.g.][]{assdg12,bfo14}, 
because the phase-space distribution of the [$\alpha$/Fe]-high and [$\alpha$/Fe]-low disk stars overlaps significantly with each other. It is difficult to distinguish these two disk components clearly using only phase-space information. Therefore, some studies suggest that thick and thin disk stars are better to be defined with their chemical properties only \citep[e.g.][]{navfa11}. 
However, this chemically defined thick and thin disks are not the same as geometrically defined thick and thin disks. To avoid the confusion, we here call this chemically defined thick and thin disks "$[\alpha$/Fe]-high" and "$[\alpha$/Fe]-low" disks. 

However we note there is also a shortcoming in the latter definition, and that the best way to define these components seem to be by age only \citep[see][and references therein]{kf11}. However, the difficulty in determining accurate ages for stars sampling a large volume so far precludes one to adopt this definition. The only exception to this was the very nearby volume completed sample of Hipparcos sub-giants (25 pc around the Sun) studied by \cite{kf98,kf11}. This volume-completed sample clearly shows two sequences in the [$\alpha$/Fe] diagram: a high-[Mg/Fe] sequence is occupied by stars older than $\sim$\-10 Gyr (here we consider the so-called 10 Gyr old transition stars found by Fuhrmann, as being part of the tail of the thick disk which is mostly composed by stars older than 12 Gyrs according to this very local sample), and a lower-[Mg/Fe] sequence, occupied by stars younger than 8 Gyrs, which are identified as thin disk stars. An age-gap between the thin and thick disks, as suggested by Fuhrmann's work, has to be still confirmed for larger samples, covering larger Galactic volumes.  Here is where the contribution from Asteroseismology is expected to be of greatest impact (see discussion in \citealt{cc15} and \citealt{fa16}).

Recently, the existence of either geometrically or chemically defined two disk components were challenged by \citet{brh12} 
 based on a large number of relatively low-resolution spectra of G-dwarfs from Sloan Digital Sky Survey (SDSS)/Sloan Extension for Galactic Understanding and Exploration (SEGUE) data. They took into account the observational selection function and analysed the structures of the mono-abundance populations which were defined by dividing the stellar samples depending on their location in the [$\alpha$/Fe]-[Fe/H] plane. Then, they derived the mass-weighted stellar distribution in the [$\alpha$/Fe]-[Fe/H], and found no clear distinction of [$\alpha$/Fe]-high or [$\alpha$/Fe]-low sequence. Also, the scale-heights of the mono-abundance disk populations showed the smooth transition from thicker to thinner disks as a function of [$\alpha$/Fe]. 
 
 However, not only mono-abundances do not directly reflect into mono-age distributions (Minchev et al. 2016, in prep), but  all recent higher-resolution spectroscopic surveys, with different selection functions and without pre-selection on kinematics, clearly show a  bimodal distribution in the [$\alpha$/Fe]-[Fe/H] plane \citep[e.g.][]{assdg12,bfo14}. 
In addition, the bimodal chemical sequences of [$\alpha$/Fe]-high and [$\alpha$/Fe]-low disks are observed not only in the solar neighbourhood, but also in a large range of Galactic radius \citep[e.g.][]{baboym11,acsrg14,nbbah14,mhrbdv14,hbhnb15}. 
For example, from SDSS the Apache Point Observatory Galactic Evolution Experiment (APOGEE) data, \citet{hbhnb15} showed that the $\alpha$-high sequence of the disk stars in the large radial range ($3<$R$<\sim$15 kpc ) follows remarkably similar locus in the [$\alpha$/Fe]-[Fe/H] plane.
 
At a discussion session devoted to Milky Way modelling at the 592. WE-Heraeus Seminar, "Reconstructing the Milky Way's History: Spectroscopic Surveys, Asteroseismology and Chemodynamical models", we discussed if or not there are two distinct populations in the Galactic disk. The participants of this discussion agreed that there are at least two distinct populations in the Galactic disk, based on the current observational evidences and successes in their models with (at least) two component disk models. In this paper, based on this conclusion of the discussion, we discuss how we can identify the two populations, what is the origin of the two distinct populations and how the geometrically thick disk and [$\alpha$/Fe]-high disk are related to each other. Discussion here is based on the discussion sessions. However, the selection of the topics and discussion in this paper may be influenced by the authors' view (and it is not intended as a review either due to the limited number of pages). The authors are responsible for the discussion in this paper.

\section{How to identify the two populations?}

As discussed in the previous Section, the current high-resolution spectroscopic surveys clearly show two distinct sequences of disk stars, [$\alpha$/Fe]-high and [$\alpha$/Fe]-low, in the [$\alpha$/Fe]-[Fe/H] distribution. Therefore, at first approximation, it seems a reasonable way to divide two populations of the Galactic disk based on the chemical abundance (but see previous Section for caveats). 

By dividing the sample of disk stars in $[\alpha$/Fe]-high and [$\alpha$/Fe]-low groups from the Gaia-ESO survey, \citet{mhrbdv14} showed that the [$\alpha$/Fe]-low disk has a negative radial metallicity gradient and negative vertical metallicity gradient, while the [$\alpha$/Fe]-high sample shows no radial metallicity gradient and a very shallow negative vertical metallicity gradient \cite[see also][using SEGUE data]{crmlb12}. 
\citet{hbhnb15}  analysed the Metallicity Distribution Function (MDF) of [$\alpha$/Fe]-high disk stars at different location of the Galactic disk within $3<R<15$ kpc and $|z|<2$ kpc. They show that at a fixed height the MDF of [$\alpha$/Fe]-high disk is very similar at different radii, although the peak of the MDF slightly decreases with increasing height. This  indicates that the chemical properties of the [$\alpha$/Fe]-high disk is radially well mixed, but the vertical negative metallicity gradient was not washed out. A similar result is found by RAVE, which has clearly shown, for different chemical elements, that the abundances gradients flatten with increasing distance from the mid-plane where [$\alpha$/Fe]-high stars are more dominant \citep{b13,bspjs14} \cite[see][for an interpretation of the flattening of the abundance gradients with increasing distance from the galactic mid-plane]{rck14,mcm14}.
These observational evidences suggest that the [$\alpha$/Fe]-high disk formed differently from the [$\alpha$/Fe]-low disk. Therefore, the [$\alpha$/Fe]-high and [$\alpha$/Fe]-low disks could be a logical way of identifying the two Galactic disk populations. 
  
Comparing the spectroscopically derived stellar parameters of \citet{assdg12}  with theoretical isochrones, \citet{hdmlkg13} 
analysed the stellar ages, and divided their sample of stars by chemical abundance, like [$\alpha$/Fe]-high and [$\alpha$/Fe]-low disks, and by age (we here call this division of the populations, "age-old" and "age-young" disks). In general, the [$\alpha$/Fe]-high disk stars are older than [$\alpha$/Fe]-low disk stars in agreement with previous results of  \cite{kf11} 
\cite[see also][for the implication from additional carbon and nitrogen abundances]{mg15}. 
 They also showed that the age-old disk stars show a clear sequence of decreasing [Fe/H] and increasing [$\alpha$/Fe] with the age. On the other hand, the age-young disk stars show a flat age-[Fe/H] relation with a large scatter and a much shallower slope of increasing [$\alpha$/Fe] with the age. Although this result can readily be explained on pure chemodynamic grounds (where only for very high-[$\alpha$/Fe] stars, there is a clear relation between age-and-[$\alpha$/Fe], whereas for lower [$\alpha$/Fe] stars the different dependencies on time in different radial bins are mixed due to the mosaic of different birth radii at each bin - see Minchev, Chiappini \& Martig, this volume), it also suggests that the age-old disk formed differently from the age-young disk. Although the age of stars is difficult to be obtained, the age-old and age-young disks would be most probably the best choice of identifying different Galactic disk components, as discussed in Section~1.
  
How the [$\alpha$/Fe]-high disk is related to the age-old disk is still not clear because of the lack of large samples with accurate age estimates. The age is obviously key information for Galactic archaeology, and will provide us chronological order of the built-up process of the Galactic disk. Therefore more accurate age estimates with the asteroseismology analysis for giant stars would be a crucial information to answer this question. However, the pioneering works based on CoRoT and Kepler light curve data revealed even more complicated situation. \citet{carmm15,mrahm15} 
found young [$\alpha$/Fe]-high disk stars with higher age accuracy than the previous isochrone based age estimates. It is important to further investigate the fraction of such "anomalies" and their spatial and kinematical properties to identify their origin and the contribution to the overall structure of the two distinct disk populations. Part of these objects could be blue-stragglers and more data is necessary to confirm this possibility.
   
\section{What is the origin of the two populations?}

 What mechanism made them so distinct? There must be some drastic change of the formation process of the Galactic disk in the past. There are three main approaches currently applied to study this issue, as recently summarized by \citet{sb15} and Minchev, Chiappini and Martig (this volume). First approach is to fit the observational data with a Galaxy model, e.g. mass model, stellar population model or dynamical model, which can characterise the properties of each components (see Binney, Girardi, Robin and Sharma, this volume). Second approach is a semi-analytic model which follows pure chemical or chemo-dynamical evolution analytically, and can associate the observed chemical signatures with the star formation history (\citealt[e.g.][]{mf89,cc97,bp99}, as well as the chemodynamical model of \citealt{kpa13,kpa15}, which adopts parametrized radial mixing prescriptions based on N-body simulations). The third approach is an ab initio approach, like cosmological simulations, which are still difficult to adjust the models to match the observational data, compared to the other approaches, but they are useful to associate the current observed properties with the cosmological formation history of the Galactic disk. 

Within the third approach there are different alternatives.  On one hand, one can have fully self-consistent cosmological simulations including star formation and chemistry which, however, still lack the required resolution in order to account for key sources of stellar radial migration, such as the galactic bar and spiral arms, (and the simulations also still prone to suffer from sub-grid physics uncertainties). On the other hand, one can use the new hybrid approach proposed by Minchev, Chiappini and Martig (2013, hereafter MCM13), in which a high-resolution simulation in the cosmological context is coupled with a pure chemical evolution model of the thin disk.
 
 An example of the first approach is the spatial structure analysis of mono-abundance populations in \citet{brlhbl12}. Most recent study of \citet{brsnhsb16} based on the APOGEE data presented that the [$\alpha$/Fe]-high disk shows a constant scale-length irrespective of [Fe/H], and the scale-length of the [$\alpha$/Fe]-high disk is smaller than the [$\alpha$/Fe]-low disk. They fitted the [$\alpha$/Fe]-low disks with different [Fe/H] by a density profile where the density increases with radius within $R<R_{\rm peak}$ and decreases at $R>R_{\rm peak}$, and found that $R_{\rm peak}$ is larger for lower [Fe/H] population of the [$\alpha$/Fe]-low disk. 
 
 Another example of the first approach is \citet{rrfcrm14} and Robin's talk at this conference who fitted the SDSS and 2MASS observational data with a sophisticated population synthesis model, the Besan\c{c}on Galaxy model. They presented that the older age-old disk (in their population synthesis model, the thick and thin disks are defined by the age, and therefore we call their thick disk age-old disk) shows a larger scale-height and a larger scale-length than the younger age-old disk. Hence, they suggested that the age-old disk formed upside-down and outside-in. On the other hand, the age-young disk has the younger disk having a larger scale-length, and shows the inside-out formation. Interestingly, the scale-length of the younger age-old disk is similar to the oldest age-young disk, $h_R\sim2$ kpc, which may indicate that the age-young population formed from the left-over gas of the age-old populations \citep[see also][]{bt11} in the inner disk. This trend was also tentatively seen in \citet{brlhbl12}. 
 
 More self-consistent method of the first approach is for example \citet{sb15} and Binney's talk at this conference. \citet{sb15} introduced a method to fit the observational data with the distribution function (they named extended distribution function, EDF) analytically coupled with [Fe/H] which is a function of age and their formation radius, including the mixing effect of the radial migration. \citet{sb15} successfully fitted the Geneva-Copenhagen Survey data applying different EDFs to the two disk components which they named thick and thin disks. This method is a promising approach to describe the chemo-dynamical properties of the distinct populations, and flexible enough to associate the EDF coupled with a certain Galactic disk formation scenario, as \citet{sb15} assumed a chemo-dynamical evolution scenario with radial migration of \citet{sb09a}. Therefore, it will help to link with the second and third approaches. 
 
One of the first models trying to explain the discontinuity in the [$\alpha$/Fe] plane by two distinct populations, within the second approach (i.e. semi-analytical chemical evolution model), is the two-infall model of \cite{cc97,cmr01}. To create the [$\alpha$/Fe]-high and [$\alpha$/Fe]-low disks, \citet{cmr01} considered the early intense star formation for [$\alpha$/Fe]-high disk formation followed by a brief cessation of the star formation, which leads to lower [Fe/H] of the inter-stellar medium (ISM) due to the fresh gas accretion. Then, the Galaxy built up new sequence of [$\alpha$/Fe]-low disks from lower [Fe/H]. \citet{cc09} computed a similar model, but in which the two components (thick and thin disk) evolve completely independently of each other. This model was shown to be able to explain the differences found among the abundance ratios measured for kinematically selected thick and thin disks by Feltzing and collaborators. According to \citet{cc09} the [$\alpha$/Fe]-high disk stars can be created as long as they have much shorter star formation period with much higher star formation efficiency than the [$\alpha$/Fe]-low disk. This is partly why the continuous star formation model in \citet{sb09b} could explain (apparent) [$\alpha$/Fe]-high and [$\alpha$/Fe]-low sequence in the solar neighbourhood. \citet{sb09b} claimed however that [$\alpha$/Fe]-high populations were created in the inner region where the star formation efficiency was higher and brought to the solar neighbourhood with radial migration, and that radial migration alone could account for the observed discontinuity in the [$\alpha$/Fe]-[Fe/H] plane. However, as shown by MCM13, radial migration alone is not able to create the observed discontinuity in the chemical plane. In addition mergers are clearly needed in order to create a thick disk with scale heights compatible with the one observed for the Milky Way (see below).
  
Based on semi-analytical chemical evolution model, \citet{bt11} discussed three possible mechanisms to build up the [$\alpha$/Fe]-high and [$\alpha$/Fe]-low sequences. First model is a continuous star formation model \citep[see Fig. 15 of][]{bt11}. This model considers that the early intense star formation built up the [$\alpha$/Fe]-high sequence up to around solar [Fe/H]. Then, from the leftover gas with almost solar [$\alpha$/Fe] and [Fe/H] immediately after the intense [$\alpha$/Fe]-high disk formation, the smooth gas accretion and gradually increasing low-level star formation built up a sequence of decreasing [Fe/H] and increasing [$\alpha$/Fe] until the star formation rate reach their peak. Because Type Ia supernovae already started contributing to the chemical enrichment in this epoch, this sequence is lower [$\alpha$/Fe] than the [$\alpha$/Fe]-high sequence. After the peak, the star formation decreases. This leads to decrease in [$\alpha$/Fe] and increase in [Fe/H] along the same [$\alpha$/Fe]-low sequence. The other two models of \citet{bt11} considered a brief cessation of star formation after the [$\alpha$/Fe]-high disk formation like \citet{cmr01} due to either a lack of the gas after the rapid [$\alpha$/Fe]-high disk formation or the gas outflow because of the intense [$\alpha$/Fe]-high disk star formation. These scenarios could be linked with the formation history of the disk in the cosmological context.  

 Third approach of cosmological simulation of the disk galaxies seems to converge to one scenario \citep[e.g.][]{bkgf04b,bsgkh12,bkwgcmm13,mcm13,sbrbr13}.  \citet{bkgf04b} serendipitously found from N-body/smoothed particle hydrodynamics (SPH) simulations that a cold dark matter (CDM)-based hierarchical clustering galaxy formation naturally create kinematically hot and geometrically thick disk at high redshift when there were multiple mergers of building blocks taking in place. Because these building blocks were tiny galaxies and therefore gas rich, the stars mainly formed in-situ in the central gas-rich disk after they merged. This would be an ideal condition for short-timescale intense star formation to build up the [$\alpha$/Fe]-high disk stars. In addition, because of the mergers, the central gas disk was kinematically hot. As a result, the stars formed from the gas disk was geometrically thick disk. Once the mergers stopped at a later epoch, much less intense, but more normal star formation supplied by a continuous low-level of gas accretion built up the [$\alpha$/Fe]-low disk component. The observed well-mixed metal abundances for the [$\alpha$/Fe]-high disk are naturally produced in the high-z gas-rich merger driven in-situ [$\alpha$/Fe]-high disk population formation as shown in \citet{bgmk05} and \citet{bt11}.
 
Related to the the third approach, more controlled numerical simulations of the disk formation are also valuable to study the formation scenario of the Galactic disk. Using a monolithic collapse model in an already formed Galactic dark matter halo, \citet{mn98} suggested that the age-old and geometrically thick disk can be formed by the kinematical scattering from giant clumpy star forming regions in the gas-rich disk at high redshift. Using a high resolution numerical simulation of a monolithic collapse model,  \citet{lrdiqw11} suggested that the radial migration can produce the geometrically thick disk. However, the simulations starting with an already established geometrically thin disk without a collapse suggested that the radial migration cannot thicken the disk \citep{mfqdms12}, which may indicate that the age-old disk should be geometrically thicker when they formed \citep[see also][]{abs16} as seen in the cosmological simulations \cite[e.g.][and MCM13]{bkgf04b,b09a,b09b}.

Finally, in the MCM13 approach, although the oldest stars do resemble the thick disk in many respects, a discontinuity in the chemical [$\alpha$/Fe]-[Fe/H] plane is not created (see also a discussion in \citealt{cc15,fa16}). Indeed, one of the goals of the MCM13 approach was to test if, once radial mixing and merger were at work, one would obtain the observed discontinuity in the chemical plane without the need of invoking a discrete thick disk (see Minchev, Chiappini, Martig - this volume). As expected, the answer was no, which suggests either the need for a discrete thick disk or for a better modelling of the inner galaxy regions (involving a complex chemical evolution in the bulge/bar/inner disk region). With the new constraints brought recently by surveys, such as APOGEE, more tight constraints on the models in the inner parts of the Milky Way will be possible.

 
In summary, these approaches are complementary and informative to each other. These approaches also enable us to link the observations of the disk galaxies at different redshift to the current properties of the Galactic disks. For example, the adaptive optics integral-fieald-spectrograph studies revealed the metallicity gradient of the lensed disk galaxies at $z\sim1-2$ where is likely a formation epoch of the [$\alpha$/Fe]-high and age-old disk. \citet{yksrl11} found a steep radial metallicity gradient of $-0.16\pm0.02$ dex kpc$^{-1}$ for a disk galaxy at $z=1.49$. If the [$\alpha$/Fe]-high disk of the Milky Way had this steep metallicity gradient, there must be some mechanism to flatten the metallicity gradient (without heating the [$\alpha$/Fe]-low disk) to explain the flat radial metallicity gradient of the Milky Way's [$\alpha$/Fe]-high disk. However, the observations of more sample of high-z disk galaxies showed the variety of the metallicity gradients, including a significant fraction of a flat metallicity gradient \citep{ljesrza15}. The evolution of the metallicity gradients should provide strong constraints on the disk formation scenarios \citep[e.g.][]{gpbsb13}, and therefore it is important to compare the metallicity gradient for the different age populations of the Milky Way with the metallicity gradient of the Milky Way-like disk galaxies at different redshifts. This seems to be finally possible to be done for the Milky Way thanks to the combination of spetroscopy and asteroseismology as recently shown by Anders et al. (submitted), who used a sample of CoRoT stars for which APOGEE spectra were taken. This turns out to be a key constraint to chemodynamical models of our Galaxy. 

\section{Geometrically vs. Chemically defined Two disk populations}

\citet{rck14} analyzed both the radial metallicity ([Fe/H]) and [$\alpha$/Fe] gradients for a cosmologically simulated disk galaxy. They found that the radial metallicity gradient at the disk plane showed the negative metallicity gradient, while the radial metallicity gradient was positive at the high vertical height, $2<|z|<3$~kpc. This trend is consistent with the observed radial metallicity gradients at the different vertical height in the Milky Way \citep[e.g.][]{ccz12,acsrg14}. 
\citet{rck14} found the old compact thicker disk and younger thinner larger but flaring disk are responsible for this trend.
Since the older disk populations are more metal poor than the younger disk population, this can drive a positive radial metallicity gradient at the high vertical height. \citet{rck14} predicted that if this is true, we should see a negative radial [$\alpha$/Fe] gradient at the high vertical height, which has been later confirmed by the observations \citep[e.g.][]{acsrg14}.
\citet{rck14} also predicted that there should be a negative radial age gradient for the disk stars at the high vertical hight, which should be able to be tested in the Milky Way as well as the external edge-on galaxies \citep[see also][]{mmssdjs15}. 
  

As mentioned in Section 1, \citet{jibls08} suggested that the scale length of the geometrically thick disk of the Milky Way  is larger than that of the geometrically thin disk. This may sound contradictory to the above discussion of a compact [$\alpha$/Fe]-high or age-old disk population and a larger [$\alpha$/Fe]-low or age-young disk population. However, \citet{jibls08} studied geometrically thick and thin disks which are defined purely by spatial distribution of stars, but not by age or chemical abundances. \citet{mmssdjs15} discussed that even if the chemically decomposed [$\alpha$/Fe]-high disk population is more compact than the [$\alpha$/Fe]-low disk population, the flaring [$\alpha$/Fe]-low disk population can contribute to the geometrically thick disk at outer radii, and lead to a larger geometrically thick disk than the [$\alpha$/Fe]-high disk. In addition, the overall stellar distribution does not have to show a clear flaring, because the scale-height of the geometrically thick disk structure is determined by the mixture of the thin and thick disk populations. The flaring [$\alpha$/Fe]-low or age-young population would be clearly identified, only when the populations are decomposed by the metal abundances or age. 

\section{Future Prospects}

 In this paper, we consider that there are at least two distinct populations in the Galactic disk stars. Currently, it seems the most sensible and also practical way to identify these populations is using the chemical properties and separate the stars to the [$\alpha$/Fe]-high and [$\alpha$/Fe]-low disks. However, there are shortcomings also related to this definition. If the accurate stellar age becomes available, the age may be a better way to define two distinct populations. Still, it depends on how the two distinct populations were established. If in the history of the Milky Way formation, something happened and affected the whole Galactic disk, e.g. the gas was expelled by strong feedback and star formation stopped in the whole disk, or there was a final gas rich merger and the gas accretion rate dropped suddenly after that. Then, the age would be the most meaningful way to distinguish the two populations. However, for example, if there was a period when the [$\alpha$/Fe]-low disk started forming in the inner region of the deep potential during the [$\alpha$/Fe]-high disk is still forming globally as seen in a cosmological simulation of \citet{bsgkh12}. Then, the age distribution of the [$\alpha$/Fe]-high and [$\alpha$/Fe]-low disks may have a significant overlap depending on the region of the disk. Therefore, both age and metal abundances of stars are important to reveal the origin of the two distinct populations of the Galactic disk. 
 
ESA's Gaia mission will soon provide us 6D (or 5D for faint stars) phase space distribution of about hundred millions of giant stars which cover a large volume of the Galactic disk, which will be supplemented with the chemical abundance information from the high-resolution spectroscopic surveys, such as Gaia-ESO, APOGEE, GALAH as well as future 4MOST and WEAVE. In addition, the K2 mission with Kepler will provide the accurate ages of the giant stars in the several different directions of the Galactic disks \citep[e.g.][]{shsjl15}. Ultimately, ESA's PLATO mission will uncover the age of bright stars in the large fraction of the sky. These data will tell us the age distribution of the [$\alpha$/Fe]-high and [$\alpha$/Fe]-low disk stars at different location of the Galactic disk, which will provide us strong constraints on the formation scenario of the Galactic disk. 
 

%
\bibliographystyle{an}
%

\end{document}